\documentclass[10 pt, conference]{ieeeconf}
\IEEEoverridecommandlockouts 
\makeatletter

\let\proof\@undefined
\let\endproof\@undefined
\let\labelindent\relax
\makeatother
\usepackage{eufrak}
\usepackage{cite}
 \usepackage{dsfont}
\usepackage{amsmath}
\usepackage{hyphenat}
\usepackage{amsfonts}
\usepackage{amssymb,latexsym}
\usepackage[psamsfonts]{eucal}
\usepackage{amsthm}
\usepackage{hyperref}
\usepackage{xcolor}
\hypersetup{
    colorlinks,
    linkcolor={blue!70!black},
    citecolor={blue!70!black},
    urlcolor={blue!70!black}
}
\usepackage{graphicx}
\usepackage{epsfig}
\usepackage{epstopdf}
\usepackage{psfrag}
\usepackage{indentfirst}
\usepackage{tikz}
\usetikzlibrary{calc,patterns,decorations.pathmorphing,decorations.markings}
\usepackage{enumitem}
\usepackage{comment}
\usepackage{float}
\usepackage[font=small,skip=0pt]{caption}
\DeclareMathAlphabet{\bm}{U}{bbm}{m}{sl}

\allowdisplaybreaks

\usepackage{enumitem}

\usepackage[utf8]{inputenc}
\usepackage{graphicx}
\usepackage{BernsteinStyle}
\usepackage{latexsym,mathtools}
\usepackage[psamsfonts]{eucal}
\usepackage{fixltx2e} 
\usepackage{comment}
\usepackage{graphicx}
\usepackage{epsfig}
\usepackage{psfrag}
\usepackage{indentfirst}
\usepackage{epstopdf}
\usepackage{comment}
\usepackage{graphicx}
\usepackage[numbered]{bookmark}
\usepackage{float}
\usepackage[font=small,skip=0pt]{caption}

\allowdisplaybreaks

\newtheorem{theorem}{\indent Theorem}

\newtheorem{proposition}{\indent Proposition}

\newtheorem{lemma}{\indent Lemma}

\newtheorem{fact}{\indent Fact}

\newtheorem{conjecture}{\indent Conjecture}

\newtheorem{corollary}{\indent Corollary}

\newtheorem{definition}{\indent Definition}

\newtheorem{example}{\indent Example}

\newtheorem{experiment}{\indent Experiment}

\newcommand\xqed[1]{%
  \leavevmode\unskip\penalty9999 \hbox{}\nobreak\hfill
  \quad\hbox{#1}}
  
\newcommand\exampletriangle{\xqed{\Large$\diamond$}}

\title{Adaptive Output-Feedback Model Predictive Control\\ of Hammerstein Systems with Unknown Linear Dynamics}
\author{Mohammadreza Kamaldar \thanks{
Department of Aerospace Engineering, University of Michigan, Ann Arbor,
MI, USA.\ttfamily \{kamaldar,dsbaero\}@umich.edu}and Dennis S. Bernstein}

\begin{document}
\maketitle

\begin{abstract}
This paper considers model predictive control of Hammerstein systems, where the linear dynamics are a priori unknown and the input nonlinearity is known.
Predictive cost adaptive control (PCAC) is applied to this system using recursive least squares for online, closed-loop system identification with optimization over a receding horizon performed by quadratic programming (QP).
In order to account for the input nonlinearity, the input matrix is defined to be control dependent, and the optimization is performed iteratively.
This technique is applied to output stabilization of a chain of integrators with unknown dynamics under control saturation and deadzone input nonlinearity. 
\end{abstract}

\begin{keywords}
Adaptive model predictive control, Hammerstein system, input nonlinearity, unknown system
\end{keywords}


\section{Introduction}

By performing optimization over a future horizon, model predictive control (MPC) provides the means for controlling systems with state and control constraints
\cite{kwon2006receding,camacho2013model,eren2017model}.
In many applications, however, an accurate model of the controlled system is not available.
In this case, data-driven MPC uses a model of the system based on data collected either prior to or during closed-loop operation
\cite{allgowerMPCrecursive,markovskyDDC,markovskyDDSC,allgowerDDMPC}.

The present paper considers output-feedback control of a special class of nonlinear systems, namely, Hammerstein systems, where the dynamics are linear but the control input is subjected to a static nonlinearity, such as control-magnitude saturation \cite{satbiblio}.
In particular, we assume that the linear dynamics of the plant are \textit{a priori} unknown, whereas the input nonlinearity is known.
These assumptions are realistic in practice when the plant dynamics are subject to unknown changes, but the control hardware is designed and tested separately from the plant.

The novelty of the present paper is to combine predictive cost adaptive control (PCAC) with an iterative receding-horizon optimization technique based on a control-dependent model.
PCAC is based on recursive least squares (RLS) for online, closed-loop system identification with optimization \cite{islam2019recursive,mohseni2022recursive} over a receding horizon performed by quadratic programming (QP) \cite{islamPCAC,tamACC2021_2}.
As shown in \cite{islam2021free}, when the online, closed-loop system identification is performed in the presence of harmonic disturbances, the resulting identified model correctly predicts the frequency, amplitude, and phase of the future response, thereby facilitating the ability of MPC to perform disturbance rejection.

The present paper accounts for the input nonlinearity by using a control-dependent model that replaces the input matrix $B$ with the control-dependent matrix $B\sigma(u)/u,$ where $\sigma$ is the input nonlinearity.
This technique is used in \cite{bernstein1994nonlinear} to compensate for input nonlinearities arising in positive real plants.
In the present paper, this technique accounts for the presence of the nonlinearity within the iteration process.
In the case where $\sigma$ is a magnitude-saturation function, this technique accounts for the saturation without the need to apply a post-optimization saturation.
In the present paper, $B\sigma(u)/u$ is handled through iteration of the receding-horizon optimization, which is performed using QP \cite{kwonpearson,kwon2006receding,LiTodorov2004,TodorovLi2005}. 
Together, these techniques comprise PCAC with an iterative control-dependent coefficient (ICD-PCAC).

ICD-PCAC is demonstrated numerically by means of the well-known chain of integrators example, which has been extensively investigated under  full-state feedback 
\cite{teel1992global,lauvdal1997stabilization}
and, more recently, under output feedback
\cite{kamaldar2021dynamic}.

The contents of the paper are as follows.
Section \ref{sec_prob} presents problem formulation.
Section \ref{sec_ID} describes online identification using RLS with variable-rate forgetting, and Section \ref{sec_IOModel} presents the input-output model and its block observable canonical form.
Section \ref{sec_MPC} states the output-feedback MPC problem, and Section \ref{Sec_ICDPCAC} presents ICD-PCAC for solving the MPC problem.
Section \ref{sec_stop_warm} presents a stopping criterion and warm starting modification for reducing the computational burden of ICD-PCAC. 
Section \ref{sec_numerics} provides numerical examples with a chain of integrators dynamics subject to input nonlinearity.
Finally, Section \ref{sec_concl} presents conclusions and future research.

The following notation is used throughout the paper.
Let $x_{(i)}$ denote the $i$th component of $x\in\BBR^n.$
The symmetric matrix $P\in\BBR^{n\times n}$ is positive semidefinite (resp., positive definite) if all of its eigenvalues are nonnegative (resp., positive).

\section{Problem Formulation}
\label{sec_prob}
Consider the continuous-time system
\begin{align}
    \dot x(t) & = A x(t) + B\sigma (u(t)) + D_1 d(t),    \label{eq:x}
\\
    y(t) & = C x(t),
    \label{eq:y}
\end{align}
where, for all $t\in[0,\infty)$, $x(t)\in\BBR^n$ is the state, $y(t)\in\BBR^p$ is the output, $u(t)\in\BBR^m$ is the control,  $\sigma\colon\BBR^m\to\BBR^m$ is the known input nonlinearity such that
\begin{equation}
    \sigma(0) =0,
    \label{eq:SH0}
\end{equation}
and $d(t)\in\BBR^q$ is the unknown harmonic or constant disturbance, and $A,B,D_1,C$ are unknown real matrices of appropriate sizes.
For a harmonic disturbance, $d\colon [0,\infty)\to \BBR^q$ is given by
\begin{equation}
    d(t) = \sum_{i=1}^{n_d} d_{\rmc,i} \cos \omega_{{\rm dis},i} t + d_{\rms,i} \sin \omega_{{\rm dis},i} t,
    \label{eq:dttd}
\end{equation}
where $n_d\ge 1$, and, for all $i\in\{1,\ldots,n_d\},$ 
$\omega_{{\rm dis},i}>0$ is a disturbance frequency, 
and the vectors $d_{\rmc,i}\in\BBR^q$ and $d_{\rms,i}\in\BBR^q$ determine the amplitudes and phases of the components of the disturbance.
The output $y(t)\in\BBR^p$ is sampled and corrupted by discrete-time sensor noise to produce the measurement $y_k\in\BBR^p$, which, for all $k\ge0,$ is given by
\begin{equation}
    y_k \isdef  y(k T_\rms) + v_k,
\end{equation}
where  $T_\rms>0$ is the sample time, and $v_k \in\BBR^p$ is the sensor noise.

The objective is to design an adaptive MPC algorithm such that, for all $x_0\in\BBR^n$, $\lim_{k\to\infty}y_k=0$.

\section{Online identification}
\label{sec_ID}
Let $\hat n\ge 1$ and, for all $k\ge 0,$ let $F_{1,k},\hdots, F_{\hat n,k}\in\BBR^{p\times p}$ and $G_{1,k},\hdots, G_{\hat n,k}\in\BBR^{p\times m}$ be the coefficient matrices to be estimated using RLS.
Furthermore, let $\hat y_k\in\BBR^p$ be an estimate of $y_k$ defined  by
\begin{equation}
\hat y_k\isdef -\sum_{i=1}^{\hat n}  F_{i,k}   y_{k-i} + \sum_{i=1}^{\hat n} {G}_{i,k} \sigma(u_{k-i}),
\label{eq:yhat}
\end{equation}
where  
\begin{gather}
   y_{-\hat n}=\cdots= y_{-1}=0,\\ u_{-\hat n}=\cdots=u_{-1}=0. 
\end{gather}
Using the identity ${\rm vec} (XY) = (Y^\rmT \otimes I) {\rm vec} X,$ it follows from  \eqref{eq:yhat} that, for all $k\ge0,$ 
\begin{equation}
    \hat y_k = \phi_k \theta_k,
    \label{eq:yhat_phi}
\end{equation}
where 
\begin{align}
     \theta_k &\isdef {\rm vec}\matl  F_{1,k}&\cdots& \mspace{-10mu}F_{\hat n,k}& \mspace{-10mu}G_{1,k}&\cdots& G_{\hat n,k}\matr\in\BBR^{\hat np(m+p)},\\
     \phi_k&\isdef \matl -  y_{k-1}^\rmT&\mspace{-5mu}\cdots&- y_{k-\hat n}^\rmT&\sigma(u_{k-1})^\rmT&\mspace{-5mu}\cdots& \sigma(u_{k-\hat n})^\rmT\matr\nn\\
     &\quad\otimes I_p \in\BBR^{p\times \hat np(m+p)}.
\label{eq:phi_kkka}
 \end{align}

To determine the update equations for $\theta_k$, for all $k\ge0$, define $e_k\colon\BBR^{\hat np(m+p)}\to\BBR^p$ by
\begin{equation}
    e_k(\bar \theta) \isdef y_k - \phi_k \bar \theta,
    \label{eq:ekkea}
\end{equation}
where $\bar \theta\in\BBR^{\hat np(m+p)}.$ 
Using  \eqref{eq:yhat_phi}, the \textit{identification error} at each step $k\ge0$ is defined by
\begin{equation}
 e_k(\theta_k)= y_k-\hat y_k.  
\end{equation}
For all $k\ge0$, the RLS cumulative cost $J_k\colon\BBR^{\hat np(m+p)}\to[0,\infty)$ is defined by \cite{islam2019recursive}
\begin{equation}
J_k(\bar \theta) \isdef \sum_{i=0}^k \frac{\rho_i}{\rho_k} e_i^\rmT(\bar \theta) e_i(\bar \theta) + \frac{1}{\rho_k} (\bar\theta -\theta_0)^\rmT \Psi_0^{-1}(\bar\theta-\theta_0),
\label{Jkdefn}
\end{equation}
where $\Psi_0\in\BBR^{\hat np(m+p)\times \hat np(m+p)}$ is  positive definite, $\theta_0\in\BBR^{\hat n p(m+p)}$ is the initial estimate of the coefficient vector, and, for all $i\ge0,$
\begin{equation}
  \rho_i \isdef \prod_{j=0}^i \lambda_j^{-1}.  
\end{equation}

For all $j\ge0$, the parameter $\lambda_j\in(0,1]$ is the forgetting factor defined by $\lambda_j\isdef\beta_j^{-1}$, where
\begin{equation}
 \beta_j \isdef \begin{cases}
     1, & j<\tau_\rmd,\\
     1 + \zeta g(e_{j-\tau_\rmd}(\theta_{j-\tau_\rmd}),\hdots,e_j(\theta_j)) \nn\\
     \quad\cdot\textbf{1}\big(g(e_{j-\tau_\rmd}(\theta_{j-\tau_\rmd}),\hdots,e_j(\theta_j))\big),& j\ge \tau_\rmd,
 \end{cases}    
\end{equation}
and  $\tau_\rmd> p$, $\zeta>0$,  $\textbf{1}\colon \BBR\to\{0,1\}$ is the unit step function, and $g$ is a function of past RLS identification errors.
To determine $g$ when $p=1$, let $\tau_\rmn\in[p,\tau_\rmd)$, and let  $\sigma_{k,\tau_\rmd}^2$ and $\sigma_{k,\tau_\rmn}^2$ be the variances of past RLS prediction-error sequences $\{e_{k-\tau_\rmd}(\theta_{k-\tau_\rmd}),\hdots,e_{k}( \theta_{k})\}$ and $\{e_{k-\tau_\rmn}( \theta_{k-\tau_\rmn}),\hdots,e_{k}( \theta_{k})\}$, respectively.
In this case, $g\colon \BBR^p\times\cdots\times \BBR^p$ is defined by
\begin{equation}
    g(e_{k-\tau_\rmd}(\theta_{k-\tau_\rmd}),\hdots,e_k(\theta_k))\isdef \sqrt{\frac{\sigma_{k,\tau_\rmn}^2}{\sigma_{k,\tau_\rmd}^2}} - \sqrt{F_{\tau_\rmn,\tau_\rmd}^{\rm inv}(1-\alpha)},
    \label{eq:gg}
\end{equation}
where $\alpha\in (0,1]$ is the \textit{significance level}, and  $F^{\rm inv}_{\tau_\rmn,\tau_\rmd}(x)$ is the inverse cumulative distribution function of the F-distribution with degrees of freedom $\tau_\rmn$ and $\tau_\rmd.$
Note that \eqref{eq:gg} enables forgetting when $\sigma_{\tau_\rmn}^2$ is statistically larger than $\sigma_{\tau_\rmd}^2.$
Moreover, larger values of the significance level $\alpha$ cause the level of forgetting to be more sensitive to changes in the ratio of $\sigma_{\tau_\rmn}^2$ to $\sigma_{\tau_\rmd}^2$.

When $p> 1,$ instead of variances $\sigma_{k,\tau_\rmd}$ and $\sigma_{k,\tau_\rmn}$, we consider covariance matrices $\Sigma_{k,\tau_\rmd}$ and $\Sigma_{k,\tau_\rmn}$, and thus the product  $\Sigma_{k,\tau_\rmn}\Sigma_{k,\tau_\rmd}^{-1}$ replaces the ratio $\sigma_{k,\tau_\rmn}^2/\sigma_{k,\tau_\rmd}^2$. 
In this case, $g\colon \BBR^{p}\times \cdots\times \BBR^{p} $ is defined by 
\begin{align}
    g(e_{k-\tau_\rmd}(\theta_{k-\tau_\rmd}),\hdots,e_k(\theta_k)) &\isdef \sqrt{\frac{\tau_\rmn}{c\tau_\rmd}{\rm tr} \big(\Sigma_{k,\tau_\rmn}\Sigma_{k,\tau_\rmd}^{-1}\big)}\nn\\
    &\quad - \sqrt{F^{\rm inv}_{p\tau_\rmn,b}(1-\alpha)},
\end{align}
where 
\begin{gather}
    a\isdef \frac{(\tau_\rmn + \tau_\rmd - p -1)(\tau_\rmd-1)}{(\tau_\rmd-p-3)(\tau_\rmd - p)},\\
    b\isdef 4 + \frac{p\tau_\rmn +2 }{a-1},\quad c\isdef \frac{p\tau_\rmn(b-2)}{b(\tau_\rmd-p-1)}.
\end{gather}

Finally, for all $k\ge0$, the unique global minimizer of $J_k$ is given by \cite{islam2019recursive}
\begin{equation}
    \theta_{k+1} = \theta_k +\Psi_{k+1} \phi_k^\rmT (y_k - \phi_k \theta_k),
\end{equation}
where 
\begin{align}
  \Psi_{k+1} &\isdef  \beta_k \Psi_k - \beta_k \Psi_k \phi_k^\rmT (\tfrac{1}{\beta_k}I_p + \phi_k  \Psi_k \phi_k^\rmT)^{-1} \phi_k  \Psi_k,
\end{align}
and $\Psi_0$ is the performance-regularization weighting in \eqref{Jkdefn}.
Additional details concerning RLS with forgetting based on the F-distribution are given in \cite{mohseni2022recursive}.

\section{Input-Output Model and the Block Observable Canonical Form}
\label{sec_IOModel}
Considering the estimate $\hat y_k$ of $y_k$ given by \eqref{eq:yhat}, it follows that, for all $k\ge0,$
\begin{equation}
y_{k} \approx -\sum_{i=1}^{\hat n}  F_{i,k}   y_{k-i} + \sum_{i=1}^{\hat n} {G}_{i,k} \sigma(u_{k-i}).
\label{eq:ykapp}
\end{equation}
Viewing \eqref{eq:ykapp} as an equality, it follows that, for all $k\ge0,$ the block observable canonical form (BOCF) state-space realization of \eqref{eq:ykapp} is given by  \cite{polderman1989state}
\begin{align}
 \eta_{k+1} &=   A_{\eta,k}  \eta_{k} +  B_{\eta,k} \sigma(u_k),\label{eq:xmssAB}\\
 y_{k} &=  C_\eta   \eta_{k},
\label{eq:yhatCxm}
\end{align}
where
\begin{gather}
 A_{\eta,k} \isdef \matl - F_{1,k+1} & I_p & \cdots & \cdots & 0_{p\times p}\\
- F_{2,k+1} & 0_{p\times p} & \ddots & & \vdots\\
\vdots & {\vdots} & \ddots & \ddots & 0_{p\times p} \\
\vdots & \vdots &  & \ddots & I_p\\
- F_{\hat n,k+1} & 0_{p\times p} & \cdots &\cdots & 0_{p\times p}
\matr\in\BBR^{\hat np\times \hat n p},\label{eq:Aetak_1}\\  B_{\eta,k}\isdef \matl  G_{1,k+1} \\
 G_{2,k+1}\\
\vdots\\
 G_{\hat n,k+1}
\matr \in\BBR^{\hat n p \times m},
\label{eq:ABBA}
\end{gather}
\begin{gather}
  C_\eta\isdef \matl I_p & 0_{p\times p} & \cdots & 0_{p\times p} \matr\in\BBR^{p\times \hat n p},
  \label{eq:cmmx}
\end{gather}
and 
\begin{equation}
    \eta_{k} \isdef \matl \eta_{k(1)}\\\vdots\\ \eta_{k(\hat n)}\matr \in\BBR^{\hat n p},
    \label{eq:xmkkx}
\end{equation}
where
\begin{align}
 \eta_{k(1)} \isdef  y_{k},
\label{eq:x1kk}
\end{align}
and, for all $j\in\{2,\ldots,\hat n\},$ $\eta_{k(j)}\in\BBR^p$ is defined by
\begin{align}
 \eta_{k(j)} &\isdef -\sum_{i=1}^{\hat n -j +1}  F_{i+j-1,k+1}  y_{k-i}\nn\\
 &\quad + \sum_{i=1}^{\hat n -j+1}  G_{i+j-1,k+1}\sigma(u_{k-i}).
 \label{eq:xnkk}
\end{align}
Note that multiplying both sides of \eqref{eq:xmssAB} by $C_\eta$ and using \eqref{eq:yhatCxm}--\eqref{eq:xnkk} implies that, for all $k\ge0,$
\begin{align}
     y_{k+1}&=C_\eta \eta_{k+1}\nn\\
     &= C_\eta(A_{\eta,k}  \eta_{k} +  B_{\eta,k} \sigma(u_k))\nn\\
    &=-F_{1,k+1}  \eta_{k(1)} + \eta_{k(2)} +G_{1,k+1} \sigma(u_k) \nn\\
    &=-F_{1,k+1}  y_{k}  -\sum_{i=1}^{\hat n -1}  F_{i+1,k+1}  y_{k-i} \nn\\
    &\quad+ \sum_{i=1}^{\hat n -1}  G_{i+1,k+1} \sigma(u_{k-i})+G_{1,k+1} \sigma(u_k) \nn\\
    & =  -\sum_{i=1}^{\hat n }  F_{i,k+1}  y_{k+1-i} + \sum_{i=1}^{\hat n }  G_{i,k+1} \sigma(u_{k+1-i}),
\end{align}
which is approximately equivalent to \eqref{eq:ykapp} with $k$ in \eqref{eq:ykapp} replaced by $k+1$.

\section{Output-Feedback Model Predictive Control Problem}
\label{sec_MPC}
Let $\ell\ge2$ be the horizon length, and, for all $j\in\{1,\ldots,\ell\}$, let $\eta_{k,j}\in\BBR^{\hat np}$    be  the computed state for step $k+j$ obtained at step $k$ using
\begin{align}
      \eta_{k,j+1} = A_{\eta,k} \eta_{k,j} + B_{\eta,k}\sigma(u_{k,j}),
      \label{eq:xfka}
\end{align}
where $\eta_{k,0}\isdef \eta_k$, $u_{k,0}\isdef u_k$, and, for all $j\in\{1,\ldots,\ell-1\}$, $u_{k,j}$ is the computed control  for  step $k+j$ obtained at step $k.$
Note that
\begin{align}
      \eta_{k,1} = A_{\eta,k} \eta_{k}+B_{\eta,k}u_{k}.
      \label{eq:xk111x}
\end{align}

Next, for all $k\ge0,$ consider the performance measure
\begin{align}
	J_k&(u_{k,1},\ldots,u_{k,\ell-1}) =  \tfrac{1}{2} \eta_{k,\ell}^\rmT Q_{k,\ell} \eta_{k,\ell}\nn\\
 &\quad +\tfrac{1}{2} \sum_{j=1}^{\ell-1} (\eta_{k,j}^\rmT Q_{k,j} \eta_{k,j} + u_{k,j}^\rmT R_{k,j} u_{k,j}) ,
 \label{Jdefn}
\end{align} 
where  
$Q_{k,\ell}\in\BBR^{\hat np\times \hat np}$ is the positive-semidefinite terminal weighting, and, for all $j\in\{1,\ldots,\ell-1\},$  
$Q_{k,j}\in\BBR^{\hat np\times \hat np}$ is the positive-semidefinite state weighting, and $R_{k,j}\in\BBR^{m\times m}$ is the positive-definite control weighting.

At each time step $k\ge0$, the objective is to find a sequence of control inputs $u_{k,1},\ldots,u_{k,\ell-1}$ such that $J_k$ is minimized subject to \eqref{eq:xfka}, \eqref{eq:xk111x}, and the constraints
\begin{gather}
    \SA v_k \le b,\\
    \SA_{\rm eq} v_k = b_{\rm eq},\\
    \underline v_s\le v_{k(s)}\le \overline v_s,~~~~~ s\in\{1,\ldots,\ell(\hat np+m)-m\},
\end{gather}
where, $\SA,\SA_{\rm eq}\in\BBR^{n_\rmc\times (\ell(\hat np+m)-m)}$, $b,b_{\rm eq} \in \BBR^{n_\rmc}$, $n_\rmc\ge0$ is the number of constraints, for all $k\ge0$,  $v_k \in\BBR^{\ell(\hat np+m)-m}$ is defined by
\begin{equation}
    v_k\isdef \matl \eta_{k,1}^\rmT & \cdots & \eta_{k,\ell}^\rmT&u_{k,1}^\rmT&\cdots&u_{k,\ell-1}^\rmT\matr^\rmT,
\end{equation}
and, for all $s\in\{1,\ldots,\ell(\hat np+m)-m\},$ $\underline v_s,\overline v_s\in\BBR$ are such that $\underline v_s<\overline v_s.$  
In accordance with  receding-horizon control, the first element $u_{k,1}$ of the sequence of computed controls is then applied  to the system at time step $k+1$, that is, for all $k\ge0,$
\begin{equation}
    u_{k+1} =   u_{k,1},
\end{equation}
and $u_{k,2},\ldots,u_{k,\ell-1}$ are discarded.
The optimization is performed beginning at step $k$ and is assumed to be completed before step $k+1.$
The optimization of \eqref{Jdefn} is performed by the iterative procedure detailed in the next section.

\section{ Iterative Control-Dependent PCAC (ICD-PCAC)}
\label{Sec_ICDPCAC}
We present an adaptive MPC  algorithm based on iteration of QP for computing $u_{k+1}$.
Let $\rho\ge 1$ denote the number of iterations, and let $i\in\{1,\ldots,\rho\}$ denote the index of the $i$th iteration at step $k.$ 
For all $k\ge0$ and all $j\in\{1,\hdots,\ell\}$, let  $\eta_{k,j|i}\in\BBR^{\hat np}$ denote  the  computed state for step $k+j$ obtained at time step $k$ and iteration $i$.
Similarly, let
$u_{k,j|i} \in\BBR^m$  denote  the computed control for step $k+j$ obtained at time step $k$  and iteration $i$. 
For all $k\ge0$, all $j\in\{0,\ldots,\ell-1\}$, and all $i\in\{1,\ldots,\rho\}$, consider the state-space prediction model
\begin{equation}
    \eta_{k,j+1|i} =   A_{\eta,k}  \eta_{k,j|i} +  B_{\eta,k} \sigma(u_{k,j|i}),
    \label{eq:eta_j1}
\end{equation}
where $A_{\eta,k}$ and $B_{\eta,k}$ are given by \eqref{eq:Aetak_1} and \eqref{eq:ABBA}, and the initial conditions are
\begin{equation}
 \eta_{k,0|i} \isdef  \eta_{k},\quad u_{k,0|i}\isdef  u_{k}. \label{eq:xmtuk1} 
\end{equation}
Note that \eqref{eq:eta_j1}--\eqref{eq:xmtuk1} implies that, for all $k\ge0$ and all $i\in\{1,\ldots,\rho\},$
\begin{align}
 \eta_{k,1|i}  &=A_{\eta,k} \eta_k + B_{\eta,k} \sigma(u_k).
\end{align}

For all $k\ge0$ and all $j \in\{ 1,\ldots,\ell-1\},$ initialize
\begin{equation}
      u_{k,j|1} \isdef u_k. 
\end{equation}
Define the control-dependent coefficient $ \SB_{k,j|i}\in\BBR^{n\times m}$ by
\begin{equation}
    \SB_{k,j|i} \isdef \begin{cases}
        \frac{B_{\eta,k} \sigma(u_{k,j|i}) u_{k,j|i}^\rmT}{\|u_{k,j|i}\|^2}, & u_{k,j|i}\ne 0,\\
        B_{\eta,k}, & u_{k,j|i}= 0,\\
    \end{cases} 
\end{equation}
which, using \eqref{eq:SH0} and  \eqref{eq:eta_j1}, implies that, for all $k\ge0$, all $j\in\{0,\ldots,\ell-1\}$, and all $i\in\{1,\ldots,\rho\}$,
\begin{equation}
    \eta_{k,j+1|i} =   A_{\eta,k}  \eta_{k,j|i} +  \SB_{k,j|i} u_{k,j|i}.
\end{equation}
For all $i\in\{2,\hdots,\rho\}$, let the computed control sequence $ \{u_{k,1|i},\ldots,u_{k,\ell-1|i}\}$  be the solution of the quadratic program
\begin{align}
    \min_{\mu_{1},\ldots,\mu_{\ell-1}} \Bigg( 
   \tfrac{1}{2} \xi_{\ell}^\rmT Q_{k,\ell} \xi_{\ell}+\tfrac{1}{2} &\sum_{j=1}^{\ell-1} ( \xi_{j}^\rmT Q_{k,j} \xi_{j} + \mu_{j}^\rmT R_{k,j} \mu_{j}) \Bigg),
  \end{align}
  \mbox{subject to:~~~} 
  \begin{gather}
\xi_{1}= \eta_{k,1|i},\\
    \xi_{j+1}= A_{\eta,k} \xi_{j}+\SB_{k,j|i-1} \mu_{j},\\
     \SA \nu \le b,\\
    \SA_{\rm eq}  \nu = b_{\rm eq},\\
    \underline v_s\le  \nu_{(s)}\le \overline v_s,~~~~~ s\in\{1,\ldots,\ell(\hat np+m)-m\},
\end{gather}
where $\nu\in\BBR^{\ell(\hat np+m)-m}$ is defined by
\begin{equation}
  \nu \isdef  \matl \xi_{1}^\rmT& \cdots & \xi_{\ell}^\rmT&\mu_{1}^\rmT&\cdots&\mu_{\ell-1}^\rmT\matr^\rmT.  
\end{equation}
Finally,  let 
\begin{equation}
	u_{k+1}=u_{k,1|\rho}.
\end{equation}

\section{Stopping Criterion and Warm Starting}
\label{sec_stop_warm}
We present a modification of ICD-PCAC that can reduce the computational burden of the algorithm.
In particular, at each step $k\ge0$, the modified ICD-PCAC uses a stopping criterion to potentially stop the iterations before reaching iteration $\rho$.
Moreover, modified ICD-PCAC uses \textit{warm starting}, that is, the control sequence obtained at the last iteration of step $k$ is used to form the control sequence for the first iteration of step $k+1.$

Let $\varepsilon>0$ be a tolerance for defining the stopping criterion.
For all $k\ge0$ and all $i\in\{1,\ldots,\rho\},$ define
\begin{equation}
    U_{k|i} \isdef \matl u_{k,1|i}^\rmT &\cdots &u_{k,\ell-1|i}^\rmT \matr^\rmT \in\BBR^{m(\ell-1)}.
\end{equation}
For each $k\ge0,$ let $\rho_k\le \rho$ be  defined by
\begin{equation}
    \rho_k\isdef \min\{\rho,\min\{i\in\{2,\ldots,\rho\}\colon \|U_{k|i}-U_{k|i-1}\|<\varepsilon\}\},
\end{equation}
and let $\rho_k$ denote the index of the last iteration ate step $k$.
Now, to do warm starting, for $k= 0$ and all $j = 1,\ldots,\ell-1,$ initialize
\begin{equation}
      u_{k,j|1} \isdef u_k, 
\end{equation}
and, for all $k\ge1$ and all $j = 1,\ldots,\ell-1,$  initialize
\begin{align}
      u_{k,j|1} &\isdef \begin{cases}
      u_{k-1,j+1|\rho_{k-1}}, & j\in\{1,\ldots,\ell-2\},\\
      u_{k-1,\ell-1|\rho_{k-1}}, & j=\ell-1.
      \end{cases} 
\end{align}

\section{Numerical Examples with Chain of Integrators}
\label{sec_numerics}
Consider the continuous-time system 
\begin{align}
    \dot {\tilde  x}(t) & = A {\tilde  x}(t) + B\sigma(u(t)) + D_1 d(t),\label{eq:xhat}\\
    \tilde  y(t) & = C   {\tilde  x}(t),
    \label{eq:yhat1}
\end{align}
where $\tilde  x\in\BBR^n$, and
\begin{gather}
    A = \matl 0_{(n-1)\times 1} & I_{n-1}\\ 0 & 0_{1\times (n-1)}\matr\in\BBR^{n\times n}, \\ B = D_1 = \matl 0_{(n-1)\times 1}\\1 \matr\in\BBR^{n\times 1},\label{eq:Cchain}
\end{gather}    
%
%
and $C\in\BBR^{1\times n}$ is arbitrary.
Note that \eqref{eq:xhat}--\eqref{eq:Cchain} represents a SISO chain of integrators with arbitrary zeros and input nonlinearity. 
Let $r\in\BBR$ be a constant command, and, for all $t\ge 0$, let 
\begin{align}
        y(t) & = \tilde  y(t) - r,\label{eq:yttt}\\
         x(t) & = \tilde  x(t)-x_*,
         \label{eq:xttt}
\end{align}
where $x_*\isdef \matl r/a_0&0_{1\times (n-1)}\matr^\rmT$.
Since $A x_*=0$ and $Cx_*=r$, using \eqref{eq:yttt} and \eqref{eq:xttt}, it follows from \eqref{eq:xhat} and \eqref{eq:yhat1} that
\begin{align}
    \dot {x}(t) & =  \dot {\tilde  x}(t) \nn\\
    & =A (x(t)+x_*) + B\sigma(u(t)) + D_1 d(t)\nn\\
    &=A x(t) + B\sigma(u(t)) + D_1 d(t),\label{eq:54xx}\\
      y(t) & = C   x(t),\label{eq:55y}
\end{align}
which are the same as \eqref{eq:x} and \eqref{eq:y}.
Thus, for the chain of integrators with arbitrary zeros and input nonlinearity given by \eqref{eq:xhat} and \eqref{eq:yhat1}, we can apply ICD-PCAC to \eqref{eq:54xx} and \eqref{eq:55y} to achieve command following as well as disturbance rejection.
In particular, note that if $\lim_{t\to\infty}x(t) = 0$, then \eqref{eq:yttt} and \eqref{eq:55y} imply that $\lim_{t\to\infty}\tilde y(t) = r$, as illustrated in Example \ref{ex:chain1}.

For all examples in this paper, we use ICD-PCAC with the stopping criteria and warm starting defined in Section \ref{sec_stop_warm}.

The input nonlinearity $\sigma$ has the property that $\sigma(u)/u$ has a removable singularity at $u=0$ with the value $\sigma(0)/0\isdef \lim_{u\to0}\sigma(u)/u.$

All examples in this paper are performed in a sampled-data control setting.
In particular, MATLAB `ode45' command is used to simulate the continuous-time, nonlinear dynamics, where the `ode45' relative and absolute tolerances are set to $10^{-5}.$
In addition, for all examples, we use MATLAB `quadprog' command to perform QP, where we choose $Q_{k,j}$ and $R_{k,j}$ to be independent of $k$ and $j$, and we thus write $Q$ and $R$.

\begin{figure}[t!]
    \centering
    \vspace{5pt}
\includegraphics{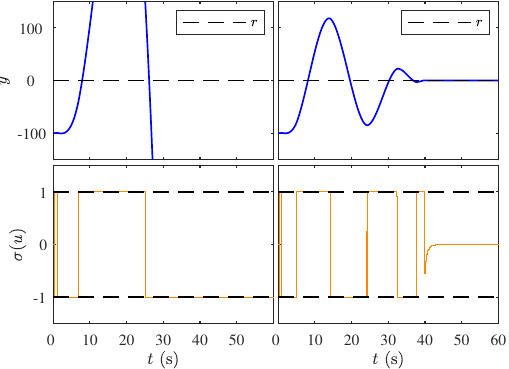}
    \caption{Example \ref{ex:adaptive_triple_integrator_Example1_sat}. Output-feedback adaptive stabilization of a chain of integrators with arbitrary zeros subject to control-magnitude saturation.
    The plots on the left show an attempt at adaptive stabilization using PCAC without using ICD coefficients,  where stabilization is not achieved. 
    The plots on the right show adaptive stabilization using ICD-PCAC.
    }
    \label{fig:adaptive_triple_integrator_Example1_sat}
\end{figure}

\begin{example}
\label{ex:adaptive_triple_integrator_Example1_sat}
Adaptive stabilization of a nonminium-phase triple integrator with control-magnitude saturation.
\rm
Consider the chain of integrators \eqref{eq:xhat}--\eqref{eq:Cchain}, where 
\begin{gather}
 n=3,\quad C=\matl -2 &-1&1\matr,\\   d(t) = 0,\quad \tilde  x(0)=\matl 50&0&0\matr^\rmT,  
\end{gather}
and $\sigma$ is the control-magnitude saturation function given by
\begin{equation}
    \sigma(u)\isdef \begin{cases} u_{\max},&u> u_{\max},\\
    u,& u_{\min}\le u \le u_{\max},\\
    u_{\min}, & u<u_{\min},\end{cases}
    \label{eq:sat1}
\end{equation}
where $u_{\min},u_{\max}\in\BBR$ are the lower and upper magnitude saturation levels.
Note that if, for all $t\ge0,$ $u(t)\in[u_{\min},u_{\max}]$, then $\sigma(u(t))=u(t)$, and the transfer function from $u$ to $\tilde y$ is given by
\begin{equation}
    G(s) = \frac{(s+1)(s-2)}{s^3},\label{eq_trip_int}
\end{equation}
which is nonminimum phase.
Let $r=0$ be the command.  Furthermore, let $T_\rms=0.1{~\rms}$, and assume
there is no sensor noise. 
Figure \ref{fig:adaptive_triple_integrator_Example1_sat} (left) shows $\tilde y$ and $u$ with BPRE without using ICD coefficients.
In this case, stabilization is not achieved.
Figure \ref{fig:adaptive_triple_integrator_Example1_sat} (right) shows adaptive stabilization using ICD-PCAC,  where 
\begin{gather}
     \hat n=3, \quad \Psi_0=10^6 I_{10},\quad \theta_0 = 0.1 \mathds{1}_{10\times 1},\\
 \tau_\rmd =80,\quad \tau_\rmn=10,\quad \alpha=10^{-2},\quad \zeta=1,\\
    \ell=200,\quad \rho=30,\quad \varepsilon=10^{-3},\\ Q={\rm diag}(10^{10},0_{4\times 4}),\quad R=1,\\ u_0=0,\quad u_{\max} =-u_{\min} = 1.\exampletriangle
\end{gather}
    
\end{example}

\begin{figure}[b!]
    \centering
    \includegraphics{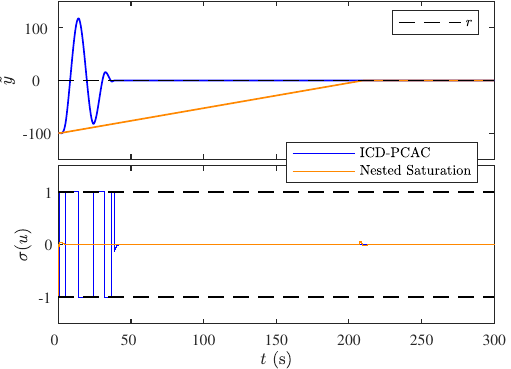}
    \caption{Example \ref{ex_nested}. Comparison of ICD-PCAC and the nested-saturation controller of \cite{kamaldar2021dynamic}. Note that ICD-PCAC is an adaptive MPC algorithm that does not require knowledge of the system, whereas the nested-saturation controller of \cite{kamaldar2021dynamic} requires knowledge of the system, and  is neither adaptive nor predictive.}
    \label{fig_nested}
\end{figure}

\begin{example}
\label{ex_nested}
Comparison of ICD-PCAC and the nested-saturation controller.
\rm
   We reconsider Example \ref{ex:adaptive_triple_integrator_Example1_sat} but we compare the performance of ICD-PCAC and the nested-saturation controller of \cite{teel1992global}.
   Since ICD-PCAC operates with output feedback, we consider the output-feedback version of \cite{teel1992global} presented in \cite{kamaldar2021dynamic}.
   Note that both \cite{teel1992global,kamaldar2021dynamic} require the knowledge of the system and are not predictive, whereas ICD-PCAC is an adaptive MPC algorithm.
   The parameters for the nested-saturation controller of \cite{kamaldar2021dynamic} are 
   \begin{gather}
       u_0=0,\quad \lambda_1=\lambda_2=\lambda_3=-0.1,\\
       \overline\varepsilon_1=-\underline\varepsilon_1= 0.24 ,\quad \overline\varepsilon_2=-\underline\varepsilon_2= 0.25 ,\\\overline\varepsilon_3=-\underline\varepsilon_3= 0.51.
   \end{gather}
   Figure \ref{fig_nested} shows $\tilde y$ and $u$, where the convergence rate is faster with ICD-PCAC than with the nested-saturation controller. 
   \exampletriangle 
\end{example}

\begin{figure}[t!]
    \centering
     \vspace{5pt}
\includegraphics{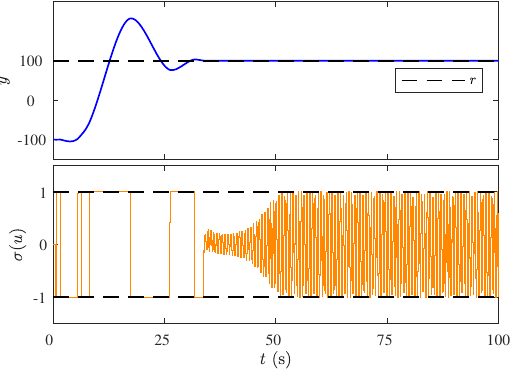}
    \caption{Example \ref{ex:chain1}. Adaptive output-feedback command following and disturbance rejection for a chain of integrators with arbitrary zeros subject to control-magnitude saturation.}
    \label{fig:adaptive_triple_integrator}
\end{figure}

\begin{example}
\label{ex:chain1}
Adaptive command following and disturbance rejection.
\rm
We reconsider Example \ref{ex:adaptive_triple_integrator_Example1_sat} but where the disturbance, the command, and the sensor noise are nonzero. In particular,
\begin{gather}
  d(t) = \sin 10t,\quad r=100,
\end{gather}
and the sensor noise $v_k$ is a zero-mean,
Gaussian white noise with standard deviation $10^{-3}.$
Figure \ref{fig:adaptive_triple_integrator} shows adaptive command following and disturbance rejection using ICD-PCAC.
    \exampletriangle
\end{example}

\begin{figure}[b!]
    \centering
    \includegraphics{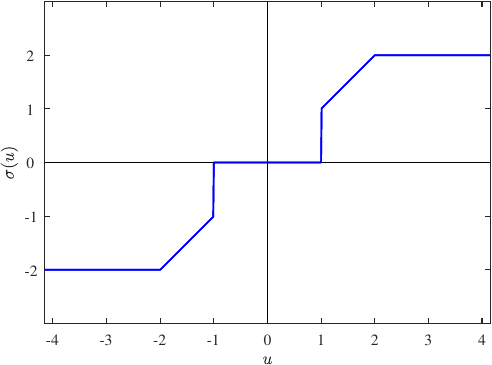}
    \caption{Example \ref{ex:dead_zone}. Control-magnitude saturation with deadzone input nonlinearity}
    \label{fig:dead_zone}
\end{figure}

\begin{figure}[t!]
    \centering
     \vspace{5pt}
\includegraphics{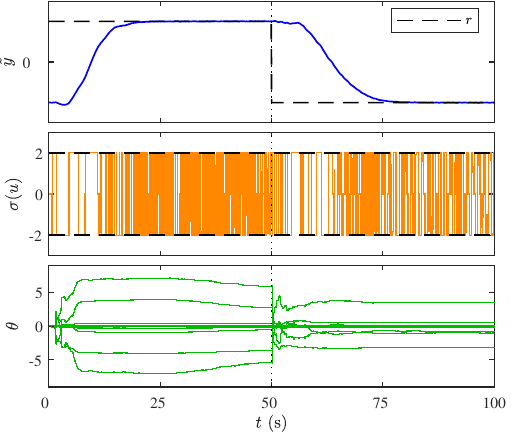}
    \caption{Example \ref{ex:dead_zone}. Adaptive output-feedback command following and disturbance rejection for a chain of integrators with arbitrary zeros subject to control-magnitude saturation with deadzone, where the command and the disturbance change abruptly at $t=50$~s.}
    \label{fig:}
\end{figure}

\begin{example}
\label{ex:dead_zone}
Adaptive stabilization with abruptly changing command and disturbance in the presence of control-magnitude saturation with deadzone.
\rm 
We reconsider Example \ref{ex:chain1}, where the command and the harmonic disturbance change abruptly at 
$t=50$~s.
In particular,
\begin{equation}
    r(t) = \begin{cases}
        100, & t\le 50{~\rm s},\vspace{1ex}\\
       -100,& t> 50{~\rm s},
        \end{cases}
\end{equation}
and
\begin{equation}
    d(t) = \begin{cases}
        \sin 10t, & t\le 50{~\rm s},\\
        2\sin 5t,& t> 50{~\rm s}.
        \end{cases}
\end{equation}
Furthermore, the input nonlinearity is control-magnitude saturation with deadzone given by
\begin{equation}
    \sigma(u)\isdef \begin{cases} 2,&u> 2,\\
    u,& 1\le |u| \le 2,\\
    0,& |u|<1,\\
    -2, & u<-2.\end{cases}
    \label{eq:sat1}
\end{equation}
Figure \ref{fig:dead_zone} shows a plot of  $\sigma(u)$ versus $u,$ and Figure \ref{fig:} shows adaptive command following and disturbance rejection using ICD-PCAC.
    \exampletriangle
\end{example}

\begin{figure}[t!]
    \centering  
   \vspace{5pt}  \includegraphics{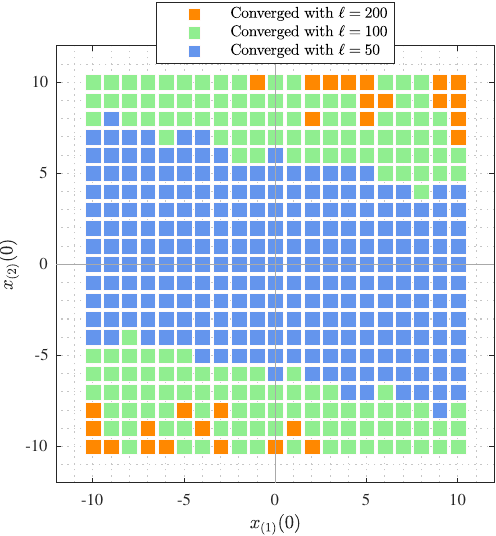}
    \caption{Example \ref{ex_daptive_triple_integrator_DOA}. Domain of attraction for the nonminimum-phase triple integrator \eqref{eq_trip_int} subject to control-magnitude saturation, where the initial conditions are $x_{(1)}(0),x_{(2)}(0)\in[-10,10]$ and $x_{(3)}(0)=0.$ Note that the domain of attraction becomes larger as the horizon $\ell$ of ICD-PCAC increases.} \label{fig_adaptive_triple_integrator_DOA}
\end{figure}

\begin{example}
\label{ex_daptive_triple_integrator_DOA}
Domain of attraction of ICD-PCAC.
\rm
We reconsider the chain of integrators in Example \ref{ex:adaptive_triple_integrator_Example1_sat}, and we investigate the domain of attraction using ICD-PCAC.
In particular, we consider the grid of initial conditions where $x_{(1)}(0),x_{(2)}(0)\in\{-10,9,\ldots,9,10\}$ and  $x_{(3)}(0) = 0.$ 
Each simulation is run for $60$~s, which since $T_\rms=0.1$~s, yields 600 step. 
The convergence criterion is $\sum_{k=580}^{600}\|x_k\|<0.01.$
Figure \ref{fig_adaptive_triple_integrator_DOA} shows the domain of attraction of ICD-PCAC for  $\ell\in\{50,100,200\}$.
%
%
Numerical simulations with larger values of $\ell$ (not shown) suggest that ICD-PCAC  provides semiglobal stabilization.
\exampletriangle
\end{example}

\section{Conclusions and Future Work}
\label{sec_concl}

The present paper considered output-feedback control of Hammerstein systems, whose linear dynamics are a priori unknown, but whose input nonlinearity is known.
To address this problem, this paper combined predictive cost adaptive control (PCAC) with an iterative receding-horizon optimization technique based on a control-dependent model.
In particular, the input nonlinearity was accounted for by using a control-dependent model that replaces the input matrix $B$ with the control-dependent matrix.
The control-dependent term is handled through iteration of the receding-horizon optimization, which is performed using quadratic programming (QP). 
ICD-PCAC was demonstrated numerically by means of several well-known examples.
For all of the numerical examples, the iteration was found to converge reliably.

Future research will focus on understanding the reasons for convergence of the iterations involving QP with the control-dependent terms.

\bibliographystyle{IEEEtran}
\bibliography{Ref,MPCrefs}

\end{document}